# Characterization of the thermo-mechanical behaviour of Hemp fibres intended for the manufacturing of high performance composites.

Vincent Placet

Department of Applied Mechanics- FEMTO-ST Institute – UMR CNRS 6174
University of Franche-Comté - 24 Chemin de l'Epitaphe. F-25000 Besançon.
e-mail : vincent.placet@univ-fcomte.fr
tél : +33 (0)3 81 66 60 55  fax : +33 (0)3 81 66 67 00

## Abstract

In this paper, the thermo-mechanical behaviour of hemp fibres (*Cannabis sativa L.*) is investigated by means of a Dynamic Mechanical Analyser. Experiments were performed at a frequency of 1 Hz, over the temperature range between 20°C and 220°C. When a periodic stress is applied to an elementary fibre, an increase in its rigidity and a decrease in its damping capacity are observed. These changes in its mechanical properties tend to stabilize after an identified number of cycles, thus providing evidence of an "adaptation" phenomenon. This specific mechanical behaviour certainly involves biochemical and/or structural modifications, such as microfibril reorientation, in the material's organisation. In addition, the behaviour of hemp fibres is affected by temperature, which acts not only as an activation factor, but also as a degradation factor with respect to the visco-elastic properties of the fibres. The rigidity and endurance of the fibres are highly affected by thermal treatment at temperatures above 150°C, and up to 180°C. Taking these results into account, polypropylene-hemp fibre composites were manufactured using a specific processing cycle. By respecting the integrity of the fibres during manufacturing, it is found that with such composites, comparatively high performance can be achieved with some specific mechanical properties. This is highly encouraging for applications requiring high mechanical performance.

## 1. Introduction

In the context of challenging environmental issues and a global energy crisis, bio-based materials are attracting increasing levels of research interest, from both academia and industry, because of their numerous advantages: renewable resource usage, low cost, biodegradability, and so on. Natural fibres such as hemp, flax and sisal have been identified as attractive candidates for the reinforcement of thermoplastic polymers. They are cheap, abundant and renewable, and have good specific properties due to their low densities. The following review papers summarise current knowledge relevant to the field of natural fibres and biocomposites: Goda and Cao [1], Yu *et al*. [2], John and Thomas [3], Baley *et al* [4]. Until recently, natural fibre reinforced polymers were used mainly in the construction and automotive industries. These markets are rapidly expanding. Other applications requiring high mechanical performance are envisaged, in order to enhance and highlight the properties of this plant-based resource. Naturally, the strength and load-bearing capabilities of composites are directly dependent on the mechanical properties of their constituents (fibres and matrix), their microstructure, and their interfacial bond strength. Some results, previously published by Bodros *et al* [5], are very encouraging and point towards the possibilities of structural applications.

Nevertheless, contrary to classical synthetic fibres (carbon, glass), the behaviour of plant-based fibres depends strictly on their temperature and humidity, as emphasised for example by Davies and Bruce [6]. Classically, during the manufacturing of composites, the process temperature is chosen in accordance with the viscosity of the polymer matrix. For polymers reinforced with natural fibres, this temperature level must not lead to a loss of the fibres' integrity. Degradation temperatures and mechanisms are relatively well known for wood fibres: from 190°C to 220°C for dry wood, depending on the species (Funaoka et al. [7], Passard and Perré [8]). Nonetheless, considerable differences exist between wood fibres and vegetable fibres stemming from annual plants, in terms of both structural and biochemical composition, and macromolecular arrangement. Literature on the topic of the thermal degradation of natural fibres is unfortunately poor. According to Gassan and Bledzki [9], the properties of jute and flax fibres are noticeably affected at temperatures of about 170°C. Xue et al. [10] show that the high temperatures (170°C to 180°C), to which fibre bundles are probably subjected during fibre processing and composite manufacturing, do not induce significant effects on their tensile properties if such temperatures are maintained for less than one hour. Mieck et al. [11] reported major damage to flax fibres after an exposure time of four minutes at a temperature above 240°C.

The main objective of the present study is to characterise the thermo-mechanical behaviour of vegetable fibres, with the aim of manufacturing composites with high mechanical performance.

The study concerns hemp fibres (Cannabis sativa L.) supplied by the LCDA company (La Chanvrière de L'Aube) in France. It should be recalled that hemp fibre bundles are composed of several tens of elementary fibres (Fig.1). Each fibre has a hierarchical structure. A fibre cell is made up of three main parts, namely the primary wall, the thick secondary wall and the lumen. The fibre cells are linked together by means of the middle lamellae. The walls consist of several layers of fibrils, which are linked together by lignins. In the secondary wall, microfibrils of cellulose are arranged in spirals and linked to hemicelluloses (Fig.1).

Initially, quasi-static tensile tests were carried out on elementary hemp fibres in order to enrich the available data, since only a limited number of studies have been carried out on the tensile behaviour of natural fibres (Baley [12, 13], Nechwatal et al. [14], Silva et al. [15], Andersons et al. [16], Kompella et al. [17]).

In addition, Dynamic Mechanical Analysis (DMA) was performed with bundles of fibres. This investigation allowed the evolution of the mechanical properties, in particular the visco-elastic properties, of the fibres to be characterised over the temperature range from 20°C to 220°C. With DMA, the sample is subjected to sinusoidal stress at a specific frequency. When the sample is scanned through a range of temperatures, its glass transition and/or other material relaxation characteristics are associated with, and can be identified by a decrease in its storage modulus, and a peak in its loss modulus and loss factor. DMA results are commonly presented in the form of a storage modulus (E'), a loss modulus (E'') and a loss factor ($\tan\delta = E''/E'$). The storage modulus describes the ability of a material to support a load, and thus represents the sample's elastic component. The loss modulus is equivalent to the sample's viscous response, and is proportional to the dissipated energy. The loss factor characterises the damping capacity of the material.

Besides this classical use of DMA, which allows the relaxation temperature of the material to be identified, harmonic tests can be used indirectly to characterize its thermal degradation. This technique has already been successfully used to study the hygrothermal degradation of wood (Placet et al. [18], [19], Assor et al. [20]). It involves the measurement of variations in the material's visco-elastic properties as a function of time, at constant temperature, and then relating these variations to the mechanical degradation of the material during the tests. It is a convenient method for determining the maximum duration and temperature level the material can be exposed to, without inducing significant irreparable damage or degradation to its properties.

Following this investigation of the mechanical properties of hemp fibres, Polypropylene – hemp fibre composites were manufactured. These were prepared using a film stacking method. The mechanical properties, i.e. Young's modulus, stress and strain at rupture, were studied as a function of the mat of the fibres.

## 2. Material and method

### 2.1 Testing of natural fibres

The hemp fibres (*Cannabis sativa L.*) tested were procured from the LCDA Company in France. They were delivered in a jumbled state. Elementary hemp fibres were isolated by hand from the initial bundles.

A DMA Bose Electroforce 3200 machine was used to perform the mechanical experiments (Fig. 2). This apparatus allows quasi-static investigations, such as harmonic tests, to be performed. It makes use of a moving magnet linear motor to apply the required forces to the sample. The applied force is measured with a 22 N load sensor with a resolution of about 10 mN, and the displacement is measured using a LVDT with a resolution of 1 μm. This apparatus is controlled by a personal computer.

For the quasi-static tests, the fibres were placed under tension at a controlled rate of displacement (0.01 $mm.s^{-1}$), until rupture. The clamping length was about 10 mm. To simplify handling of the fibres, they were positioned on a self-adhesive paper frame with a window. A drop of glue was added to each extremity and the paper frame then clamped onto the testing machine. Before the beginning of each test, the paper frame was cut. An optical system was used to assist the fixation of the fibres, to observe the fibre during the test, and to ensure that no slippage occurred at the level of the clamps. Images were recorded with a video camera equipped with a CCD detector (IXC 800, I2S), connected to an image analysis system (MATROX). A tele-objective lens (TV ZOOM LENS 18-108 mm, F 2.5) was used.

The mechanical properties, i.e; Young’s modulus, ultimate strain and stress were assessed for about 50 fibres. To determine local data (stress and strain) from global data (force and displacement), the fibres’ dimensions are needed. The diameter of the fibres was evaluated before mechanical testing, on the basis of the average of optical measurements made at nine different spots. Each fibre is considered to be a full cylinder, and its modulus is calculated from the linear part of the stress-strain curve.

For the harmonic tests, a static strain was applied to the fibres in order to maintain the samples under positive net tension throughout the experiment. A sinusoidal force was applied around this mean value of strain. Wintest software was used to pilot the DMA and compute the visco-elastic properties from the force and displacement signals, and the sample size. For each sampling interval, the length used to calculate the strain was the actual length of the sample, rather than its initial length. This “corrected sample length” enabled length variations due to thermal expansion and visco-elastic creep to be corrected for. A climatic chamber was used to control the temperature of the sample with a stability of about +/- 0.3°C. The temperature was measured with a thermocouple, placed a few millimetres from the sample.

Two types of experiment were performed using harmonic tests. The first series consisted in measuring variations of the fibres’ visco-elastic properties (storage modulus, loss modulus and loss factor) as a function of temperature. The temperature was varied between 30°C and 200°C, at a frequency of loading of 1 Hz, and the dynamic amplitude was about 2% of the ultimate stress. The fibres’ visco-elastic properties were measured at 5°C intervals, under isothermal conditions, after a stabilisation period of five minutes before each measurement. The heating rate was about 3°C per minute between each plateau temperature.

During the second series of tests, the fibres were tested at different plateau temperatures (25°C, 100°C, 150°C, 200°C and 220°C), at a frequency of 1 Hz, for a maximum duration of 16 hours. The visco-elastic properties were measured every 90 seconds. The heating rate used to reach each plateau temperature was 3°C.min$^{-1}$. For each plateau temperature, three samples were analysed in order to check the repeatability of the measurements. The dynamic amplitude applied was about 10% of the ultimate stress.

It is well known that the material history of biological materials, temperature and water content variations in particular, affect their visco-elastic properties. The fibres are therefore always dried in a steam room at 103°C before being tested.

As the well-known sample gluing method, used for quasi-static tests, is not appropriate for temperature sweep tests, a specific clamp was developed for these investigations. Indeed, since the glue can directly perturb the measurement of a sample's visco-elastic properties, only mechanical clamping solutions are acceptable. The fibres were thus positioned between two plates, one of which was machined with a groove to ensure correct centring and axial alignment of the specimen's attachment areas. The groove also limits crushing of the fibres. The screws used to clamp the two plates together were equipped with a compression spring, in order to limit the sets resulting from thermal expansion effects. Despite these adaptations, some of the elementary fibres ultimately broke at the level of the clamps. To limit such test failures, bundles of fibres were preferred to elementary fibres, thus leading to improved strength at the level of the clamps. The depth of the grooves used for the bundles was about 100 microns.

## 2.2 Composite processing

Thermoplastic / hemp fibre composites were manufactured using a film stacking process. This method consists in heating and compressing a sandwich of polymer films and fibre mats. Panels of 100 mm by 100 mm were made. Polypropylene films with a thickness of 1 mm were prepared from pellets, using an injection machine (BOY 22M). Two types of fibre mat were prepared from jumbled fibres: mats of randomly scattered fibres and mats of unidirectional fibres. For unidirectional mats, the fibres were manually carded.

The composites were moulded from a stack of three polymer films interleaved with hemp fibre mats. This stack was placed in a mould, on an electromechanical press equipped with a furnace, with which the assembly was heated to 180°C. At this plateau temperature, the stack was compressed at 10 bars for 3 minutes, following which the composite was allowed to cool to room temperature. The pressure was maintained at a constant value until 160°C (Fig. 3).

## 2.3 Characterization of the composites

Dog-bone shaped specimens (showing a reduced gauge section and enlarged shoulders) were produced from the composite plates, and then tested in accordance with the D638 ASTM standard. The gauge length was 45 mm, and the cross section around 8 x 3.5 $mm^2$. For composite panels manufactured with unidirectional fibre mats, tensile specimens were taken in two main directions: along the principal alignment axis of the fibres (0° orientation), and perpendicularly to this axis (90° orientation). For composites reinforced with randomly scattered fibres, tensile specimens were taken in random directions.

A commercial electromechanical machine (Instron 6025) was used to carry out the tensile strength tests. The samples were clamped using wedge action grips. Their deformation was measured with an extensometer. The samples were subjected to a displacement control, with a moving crosshead speed of 0.2 $mm.s^{-1}$. The displacements were measured by an optical encoder, with a resolution of 10 microns, and the tests were carried out until rupture. Since the studied materials do not have a strong linear relationship between stress and strain, even at very low stresses, a tangent method was used for the determination of Young's modulus. The experimental strain-stress curve was fitted by a polynomial regression (Davis, [21]), from which the tangent modulus, i.e. the slope, of the stress-strain curve was determined at a specific value (strain of 0.3 % ).

In parallel, several samples were taken from different parts of each panel, in order to determine the density of the composites.

# 3. Results and discussion

## 3.1 Mechanical behaviour of hemp fibres

**Quasi-static tensile tests**

Fig. 4 depicts the typical load-displacement curve recorded for elementary hemp fibres. Fifty fibres with an average diameter of about 40 μm were tested. Table 1 summarises the mechanical properties measured during this quasi-static campaign. The data collected for these fibres is compared with values taken from the literature relevant to synthetic and natural fibres. The mechanical properties, in particular Young's modulus and the ultimate stress values, are found to be lower than those published in the literature. In the case of Young's modulus, the values may have been underestimated due to approximations and some of the hypotheses used in the calculations. Indeed, it is well known that many difficulties are encountered when characterizing the mechanical properties of natural fibres. Firstly, as for all biomaterials, the properties of natural fibres are highly dependent on temperature, humidity, test duration and rate, and material heterogeneities. Furthermore, additional problems are encountered due to variations in the geometrical shape and dimensions of the fibres. The calculation of intrinsic material properties on the basis of the results of a tensile test is subject to several uncertainties: variations in the fibre diameter, influence of the clamping length, measurement of the elongation, clamping effects … (Nechwatal *et al*. [14]), and also depends on the equipment used for tensile testing, the methodology used to measure the fibre's cross-sectional area and the methodology used to compute Young's modulus (Silva et al. [15]). In order to convert global measurements (load and displacement) to local data (stress and strain), we considered the elementary fibres to be samples with a constant, cylindrical full section. However, rather than being cylindrical, natural fibres tend to be polygonal in shape, like all biological cells. Moreover, natural fibres are not full, as they have a lumen. Finally, it is important to point out that natural fibres are not constant in diameter, along their length. In the present study, we adopted a constant diameter, based on the mean of 9 measurements along the fibre. These approximations tend to underestimate the value of Young's modulus, and can also explain the large spread in the collected values observed from one fibre to another.

Concerning ultimate stress, the low measured value could be explained by the size of the defects present inside the fibre or on its surface. The MEB pictures shown in Fig. 5 illustrate the presence of such superficial flaws. Some of these significant defects are attributed to the fibre extraction process (Baley and Lamy [26]). Nechwatal *et al*. [14] highlights the influence of clamping length on the failure of fibres. The dependence of failure strength on clamping length is a well-known characteristic of all fibres. The variability of the measured tensile strength could be explained by the distribution of defects or inhomogeneities within the fibres. The longer the stressed length the higher the number of defects, present in the stressed fibre segment, contributing to weakening of its structure. Zafeiropoulos et al. [23-24] demonstrated that strength decreases with increasing clamping length. Silva and al. [15] recently reported a contradictory result for sisal fibres, and was able to clearly distinguish between the influence of the size and the number of defects. In effect, these authors clearly showed that neither tensile strength nor Young's modulus are a function of gauge length (for gauge lengths between 10 and 40 mm). Nevertheless, for all materials, the mean strength is controlled by the mean defect size, and the mean flaw size does not change with gauge length. Thus, the number of defects per unit length seems to be the judicious factor accounting for the influence of gauge length; in particular, it controls the Weibull modulus. In fact, in order to describe fibre strength, a very popular method among material scientists consists in the application of Weibull statistics. The latter have been used by several authors, to quantify the degree of variability in natural fibre strength [15-17, 24-25]. Indeed, the Weibull law is one of the most suitable and reliable means of describing the failure probability distribution of fibres. According to Weibull statistics, the probability of failure for a fibre of length L, and an applied strength $\sigma$, is given by:

$$P_f = 1 - \exp\left( -\frac{L}{L_0}\left(\frac{\sigma}{\sigma_0}\right)^m \right) \quad (1)$$

where m is the Weibull modulus, $\sigma_0$ is the characteristic strength and $L_0$ the largest fibre length containing only one flaw.

For reasons of simplicity, $L_0$ is often chosen to be unity. In this case, Eq. 1 becomes:

$$P_f = 1 - \exp\left( -L\left(\frac{\sigma}{\sigma_0}\right)^m \right) \quad (2)$$

where L is a dimensionless quantity.

Using Eq.2, after rearrangements, we can write:

$$\ln\left(\ln\left(\frac{1}{1-P}\right)\right) = m\ln(\sigma) - m\ln(\sigma_0) + \ln(L) \qquad (3)$$

A straight line, with a slope given by the Weibull modulus (m) should thus be found when $\ln(\sigma)$ is plotted as a function of $\ln\left(\ln\left(\frac{1}{1-P}\right)\right)$. Usually, the value of P is estimated using a function known as the probability index.

$$P = \frac{i - 0.5}{n} \qquad (4)$$

where n is the number of data points and i is the rank of the $i^{th}$ data point.

Although the length of the gauge has no influence on the fibre's strength, it has a major influence on the Weibull modulus. In reality, the number of flaws increases with increasing volume. Silva and al. [15] reported that the Weibull modulus decreased from 4.6 to 3 when the gauge length was increased from 10 mm to 40 mm. For this reason, and taking into account the complexity of performing tensile tests on short fibres, a moderate clamping length of 10 mm was adopted for the tests described here. Fig 6. depicts the Weibull distribution plotted for 15 elementary hemp fibres. The Weibull modulus is determined using a linear regression, and its value (m = 2.86) is in the same range as that determined by Silva et al. for sisal fibres (m = 4.6 for a gauge length of 10 mm), and by Kompella and Lambros [17] for cotton and wood fibres (respectively 1.77 and 1.26).

The relatively low value of the Weibull modulus determined for these hemp fibres can be associated with the high number of defects they contained; this can be attributed to the harshness of the fibre extraction process.

On another hand, a non-linear behaviour can be identified in the first part of the load-displacement curve (Fig. 4). It is well known that the initial portion of such curves begins with either a concave or a convex plateau, rather than a straight line. The shape of the initial portion may be influenced by numerous factors such as: seating of the test piece in the clamps, straightening of a test piece that has been initially bent by residual stresses, etc. However, in the case of natural fibres, and particularly in the case of flax fibres, Baley [13] showed that the disrupted initial portion of the load-displacement curve disappeared when the fibre was cycled with a nominal load value above the threshold load. This author proposes an explanation involving the reorientation of microfibrils during the initial stress applied to natural fibres [13]. Murherjee and Satyanarayana [22] and Silva et al. [15] hypothesised the collapse of the weak primary cell walls and delamination between cells. These theories will be further discussed in the present section.

**Harmonic tests**

In addition to quasi-static experiments, harmonic tests deliver interesting additional information concerning the rheological behaviour of hemp fibres. Variations in their mechanical properties were recorded as a function of time (or number of stress cycles). In order to avoid any uncertainties or errors resulting from the assumptions made to determine Young's modulus, a normalized modulus is considered in the following. Contrary to all expectations, periodic stress did not lead to fatigue of the material (i.e. to a decrease in its mechanical properties) but, quite to the contrary, to an increase in the fibre's rigidity. In Figure 7, it can be clearly seen that the rigidity of the fibre increases until, after 60 000 cycles, it is 1.6 times greater than its initial value at 25°C. The damping capacity of the fibre also decreases (decreasing by a factor of 2 after 60 000 cycles). This corroborates the results already revealed by Baley [12] concerning flax fibres: after 198 cycles, the Young's modulus increased to 1.67 times the value determined after the first cycle (frequency: 0.0725 Hz, stress amplitude: 60% of the ultimate stress).

It is also clear that the rigidity of the fibre increases until a constant value is reached, after a certain number of cycles, suggesting the presence of an "adaptation" phenomenon. The explanation for this complex and unforeseen mechanical behaviour certainly lies in the macromolecular organisation of the cell wall. We recall that the mechanical strength of natural fibres is mostly provided by the $S_2$ layer of the cell wall in the longitudinal direction, in particular by the crystalline cellulose microfibrils which are spirally wound in a matrix of amorphous hemicellulose and lignin. These microfibrils are tilted by an angle of the order of 10° with respect to the axis of the fibre [27]. The increase in rigidity of the fibre resulting from cyclic stresses could thus be attributed to several phenomena:

- At the level of the microfibrils, the longitudinal tension in the fibres could involve not only tensional stresses, but also torsion effects. This torsion stress, which initially impedes the tension, could subside or relax as a result of repeated stress.
- It is also probable that rearrangements and re-orientations of the cellulose microfibrils and/or changes in the cristallinity fraction occur in the fibres.
- The hypothesis of collapse of the weak primary cell walls, and delamination between cells [15, 22] can also be envisaged.

Only macromolecular investigations could allow the above hypothesis to be verified.

### 3.2 Thermal activation or thermal degradation of the mechanical properties of hemp fibres?

In the foregoing we have described the strong dependence on temperature of the mechanical properties of polymers. To evaluate the significance of temperature on their mechanical properties, fibre bundles were tested. Fig. 8 depicts the variation of the visco-elastic properties of bundles submitted to temperature ramps. A strong difference can be noticed between the heating ramp and the cooling ramp. This is due to the previously explained cycling effect. In order to isolate the influence of temperature, the fibre has to be initially pre-cycled, until its mechanical properties have stabilized. In the following analysis, only those values recorded during cooling phases are considered.

For temperatures ranging between 20°C and 200°C, it can be clearly seen that the storage modulus decreases and the loss factor increases. As for all polymer materials, these variations correspond to a material transition. They are, however, less marked and less obvious than in the case of the glassy transition of a homogeneous polymer. Contrary to the case of classical homogeneous polymers, the concept of a glass transition temperature does not apply to vegetable fibres or tissues. Rather, we are referring to *in situ* transitions of constitutive polymers or to a softening temperature, in the range in which their rheological properties vary rapidly. In fact, the cell wall of organic fibres is characterised by a complex biochemical organisation: a blend comprising three main polymers, i.e. cellulose, hemicellulose and lignin. Some of these, namely cellulose and hemicelluloses, exhibit a linear arrangement, whereas lignins are branched and constitute a cross-linked network. Moreover, hemicelluloses and lignin are completely amorphous whereas cellulose is partly crystalline. The rheological behaviour of plant-based fibres thus results from the combined influences of the intricate response of each polymeric component.

The results obtained for hemp fibres can be compared to those observed by Kelley *et al.* [27] in the case of wood (Fig. 9). In this comparison, we consider the curve for wood with a moisture content of 5%. The tan$\delta$ responses are attributed to three material transitions (Placet [19]). The parameter $\alpha_1$ is attributed to the *in situ* lignin transition, and $\alpha_2$ corresponds to the *in situ* hemicellulose transition. At low temperatures, the $\beta$ transition characterizes a secondary transition involving small-scale molecular movements only. For dry hemp fibres, the rheological behaviour is similar to that of wood. For the temperature range from 20°C to 200°C (Fig. 8), the increase in visco-elasticity of the fibre could be attributed to the succession and partial stacking of *in situ* glass transitions of the hemicelluloses and lignins. Some small differences can be noticed between wood and hemp fibres, such as: a loss factor of 0.028 at 150°C for hemp fibres, as opposed to 0.04 for spruce wood. However, it is obvious that a macroscopic wood sample cannot be directly compared to annual plant fibres, since anatomical and biochemical differences still exist.

It is thus clear that, for the considered range, temperature activates visco-elastic properties of the fibre. However, one may wonder if one part of the apparent thermal activation effect could not be induced by thermal degradation. The question to which Fig. 10 attempts to provide a response is: *"to what extent can thermal degradation alter the mechanical properties of hemp fibres ?"*

From the time varying rheological measurements shown in Fig. 10 and Table 2, significant storage modulus variations can be identified at different plateau temperatures, which can be related to distinct biochemical modifications and degradations. Fig. 10 and Table 2 clearly show that above 150°C, the fibre rigidity decreases, especially when the temperature is raised. After one hour of stress cycles at 1 Hz, the rigidity of the fibre is improved by a factor of 1.32 at 25°C, a factor of 1.2 at 150°C and a factor of 1.06 at 220°C. The endurance of the fibres is also decreased over this same range (Table 2). Dynamic mechanical analysis of the fibre at different temperatures thus allows the regions of thermal activation and thermal degradation to be mapped.

### 3.3 Mechanical performance of polypropylene-hemp fibre composites

The aforementioned results from the thermo-mechanical characterisation of hemp fibres have been used to define an experimental procedure, in particular to choose an appropriate temperature for the manufacture of a composite material (Fig.3). This temperature is a compromise between the degradation temperature of hemp fibres and the viscosity of the polymer matrix.

Table 3 summarises the average mechanical properties of polypropylene and the manufactured composites. The inclusion of randomly scattered hemp fibres in the polypropylene allows the rigidity of material samples to be doubled. Unidirectional fibre reinforcements appear to be a promising solution for the manufacture of composites with high mechanical performance. In fact, the Young’s modulus of samples cut in the PP-UD hemp boards, in the direction of the fibres, is about 12 times higher than that of polypropylene alone. This encouraging result is nonetheless mitigated by the composite’s weak failure stress. This can be directly attributed to the weak ultimate stress of the hemp fibres used, to the interfacial adhesion between the fibres and the polymer, and also to heterogeneities in the composite panels. Beyond the determination of an appropriate temperature, other manufacturing parameters need to be investigated in order to improve various mechanical properties of the composite.

## 4. Conclusion

In this paper, the thermo-mechanical behaviour of hemp fibres has been studied with the view to manufacturing high performance natural fibre-reinforced composites. The mechanical testing of fibres with small diameters is not trivial. Hemp fibres subjected to quasi-static tensile experiments and temperature-controlled harmonic tests reveal a complex behaviour, involving several mechanisms, which are sometimes antagonistic:

(i) an increase in the rigidity of the fibres during the initial stress which is certainly related to structural rearrangements in the fibre,

(ii) thermal activation of their visco-elastic properties, over the temperature range from 30°C to 200°C, corresponding to *in situ* relaxation of the constituent polymers (hemicelluloses and lignins),

(iii) a decrease in rigidity and endurance, attributed to thermal degradation of the cellular walls, at temperatures between 150°C and 180°C.

Indeed, the plant cell wall has a complex composition and organisation. The macroscopic properties of cloth made from plants can be attributed this intricate structure. To precisely determine and understand the intrinsic mechanical properties of natural fibres, the researcher needs to resort to more sophisticated devices and tools. The use of a comprehensive model and ultrastructure investigations (at micron and nanometric scales) would certainly be the most relevant solution to obtain objective and accurate results.

Nevertheless, the modest study presented here has enabled the thermal activation and degradation regions of hemp fibres to be mapped. This crucial information is indeed needed, to define a viable processing cycle for the manufacture of a composite. The polypropylene-hemp fibre composite manufactured according to these criteria clearly demonstrates the potential for obtaining high performance materials.

The next important step will be to study the composite material's behaviour under severe climatic conditions. Indeed, ageing and durability are of fundamental importance for natural fibre reinforced polymers, and have yet to be investigated.

**Acknowledgements**

The author thanks LCDA Company in France (La Chanvrière De l'Aube) for supplying hemp fibres.

a.

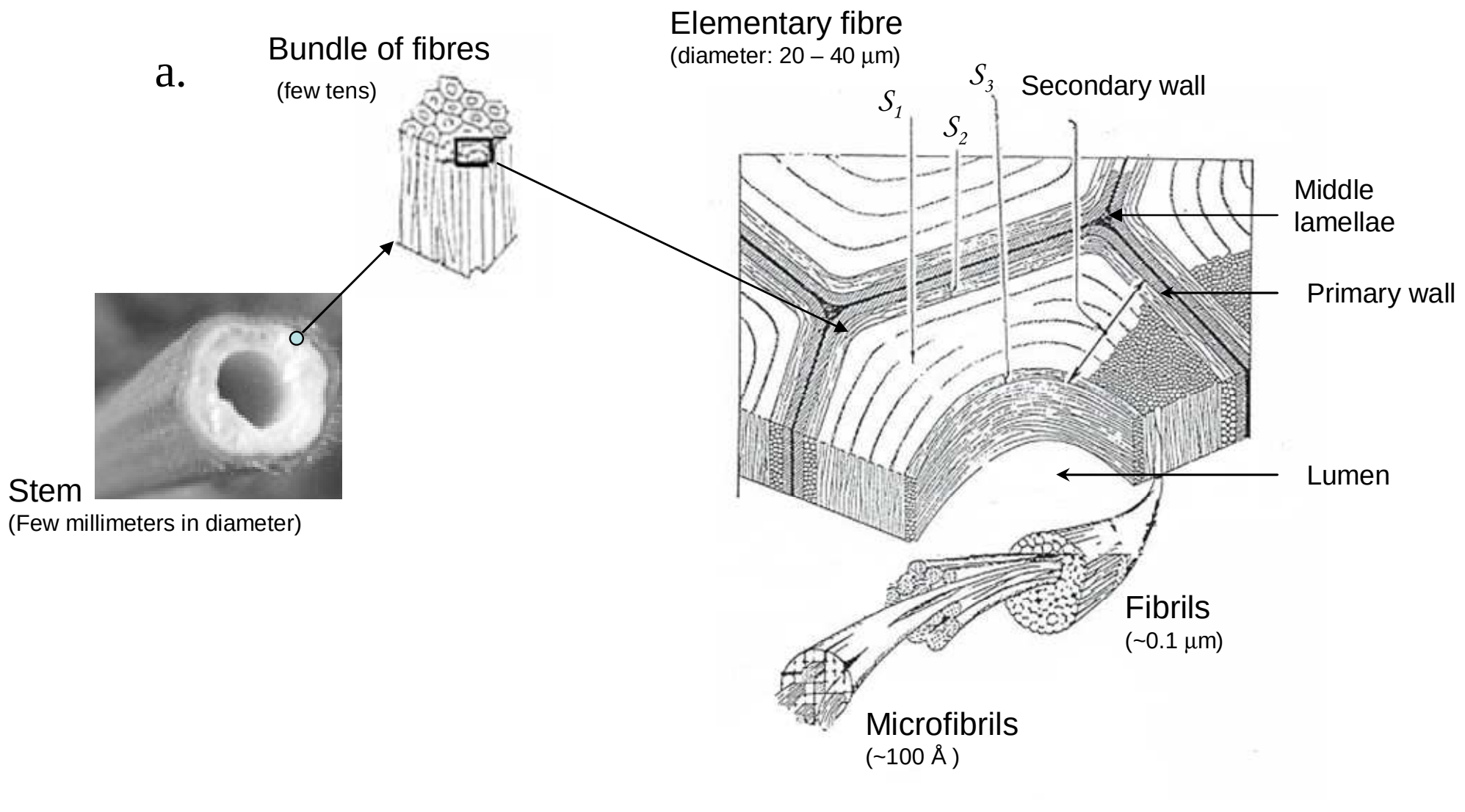

b.

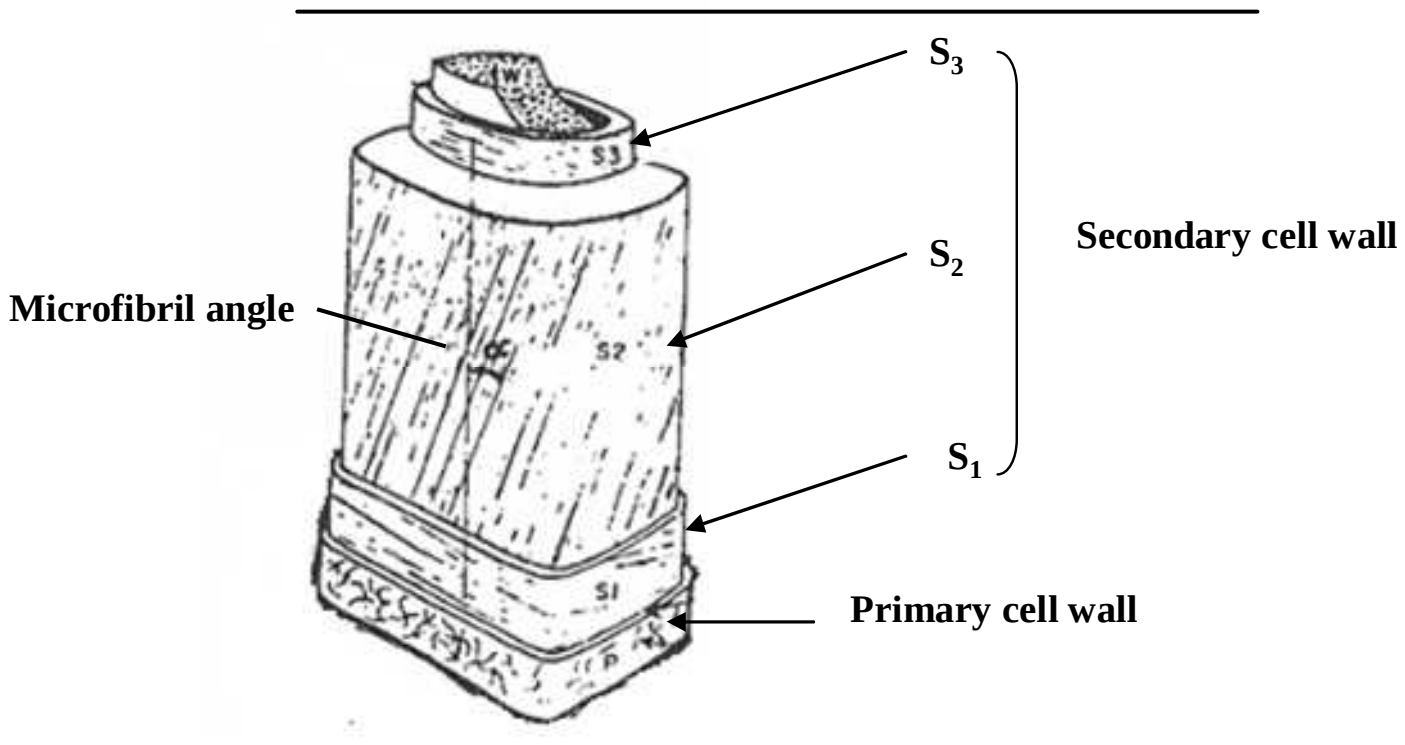

**Figure 1.a: Schematic representation of a natural fibre from stem to microfibrils (inspired from Frey-Wyssling). 1.b. Schematic structure of a natural fibre cell wall**

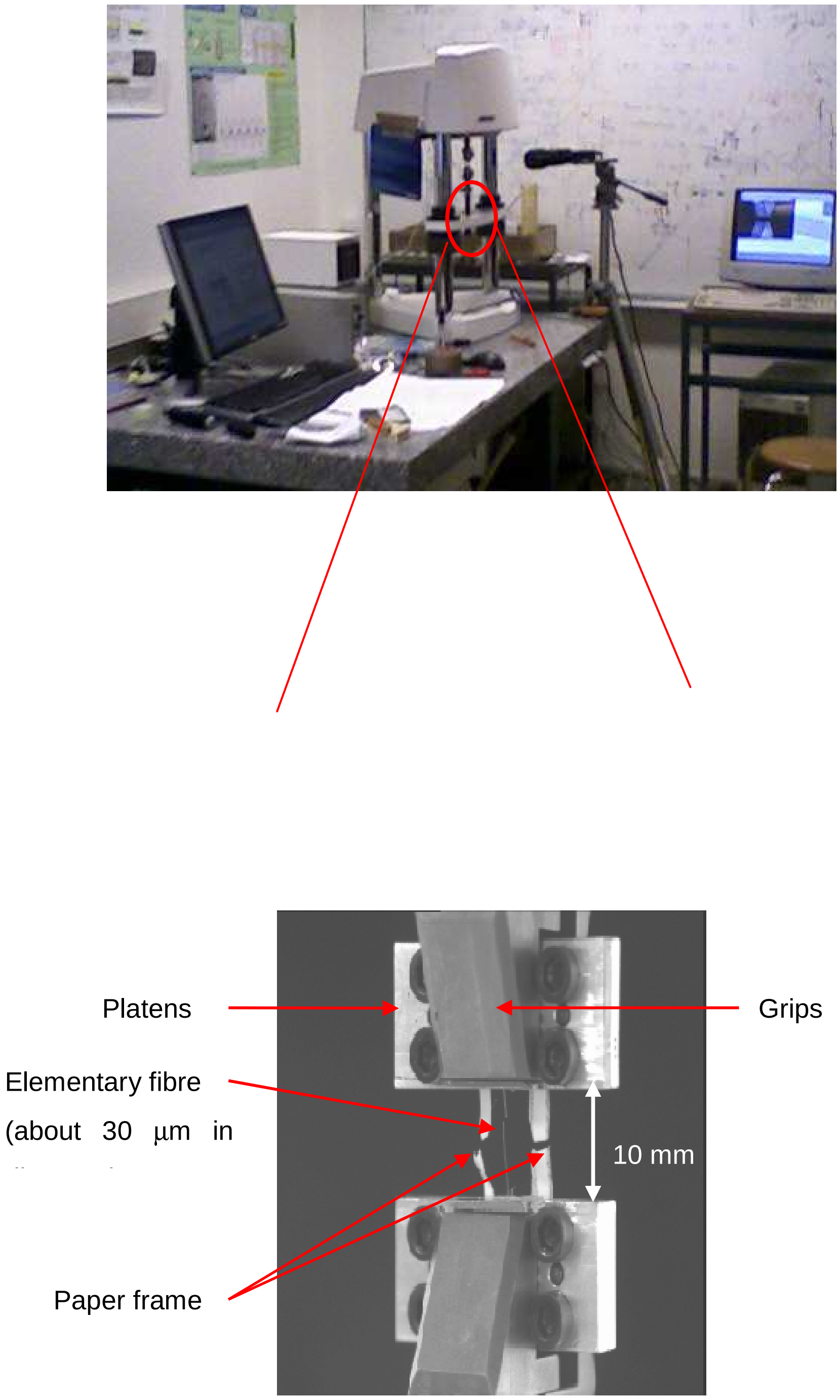

**Figure 2: DMA setup – Fibre clamping**

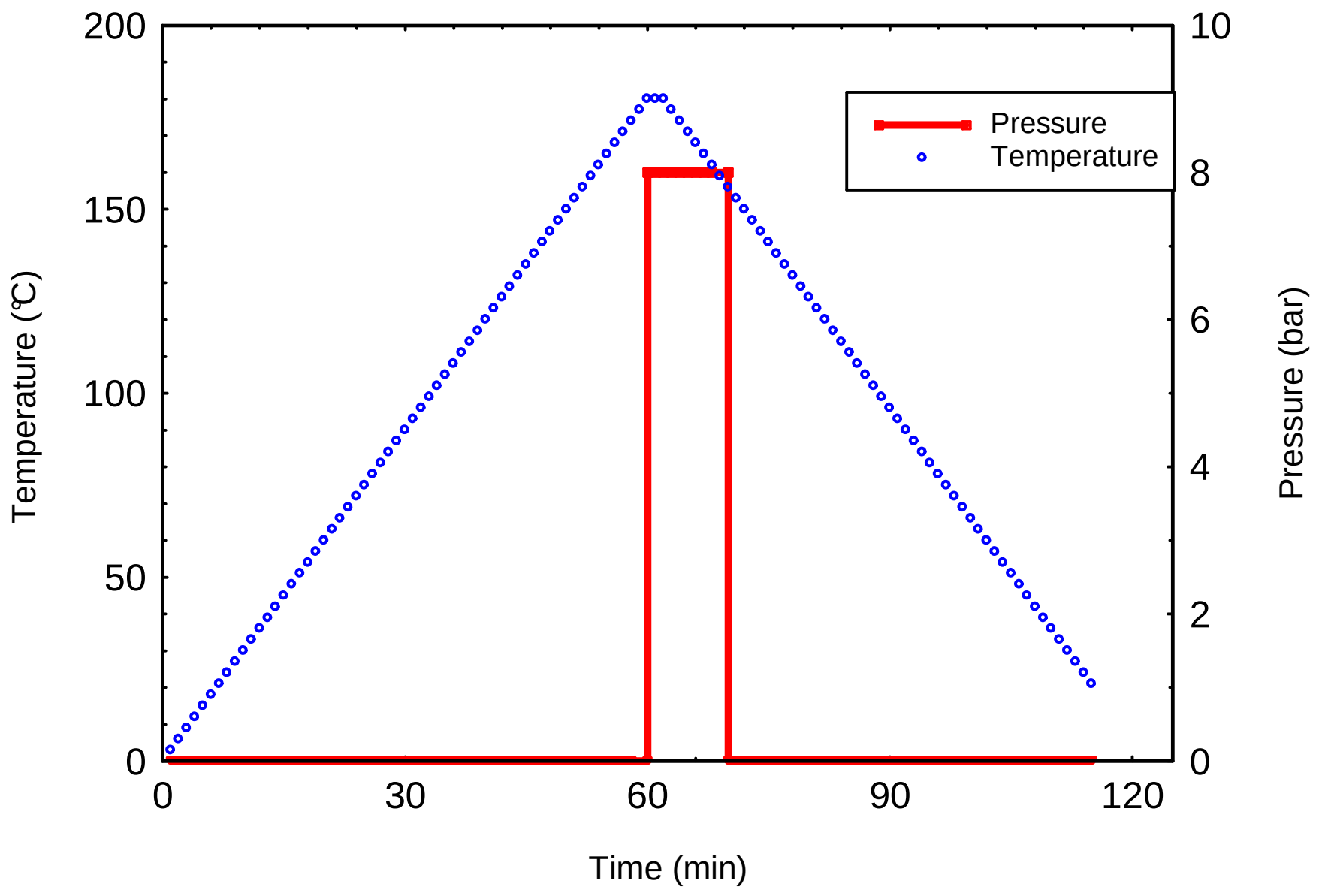

**Figure 3: Processing cycle of Polypropylene-hemp fibres composite – Temperature and pressure versus time.**

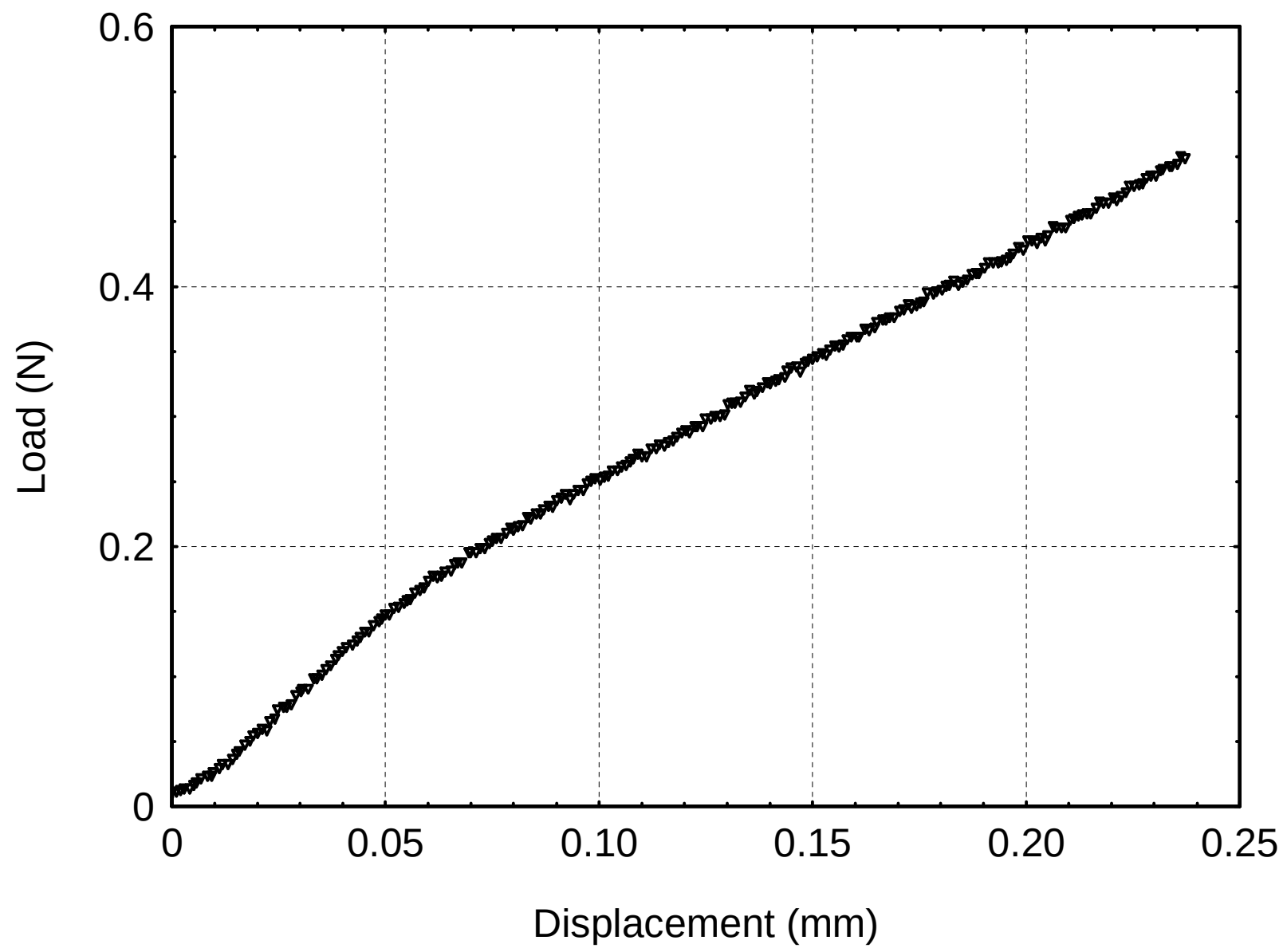

**Figure 4 : Typical load-displacement curve for hemp fibre. Displacement speed: 0,01 mm.s$^{-1}$. Gauge length: 10 mm.**

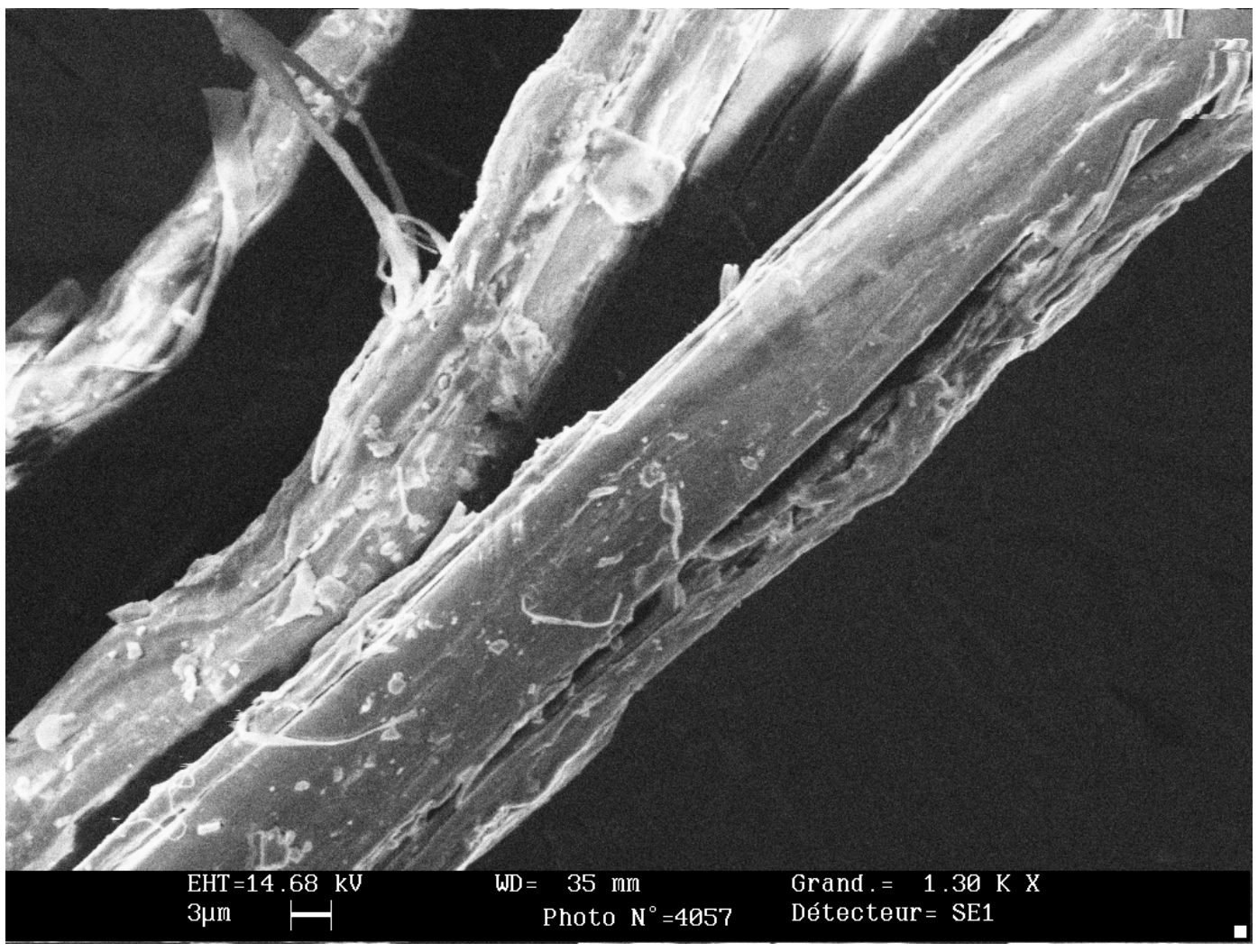

**Figure 5: Bundle of hemp fibres. Examples of defects.**

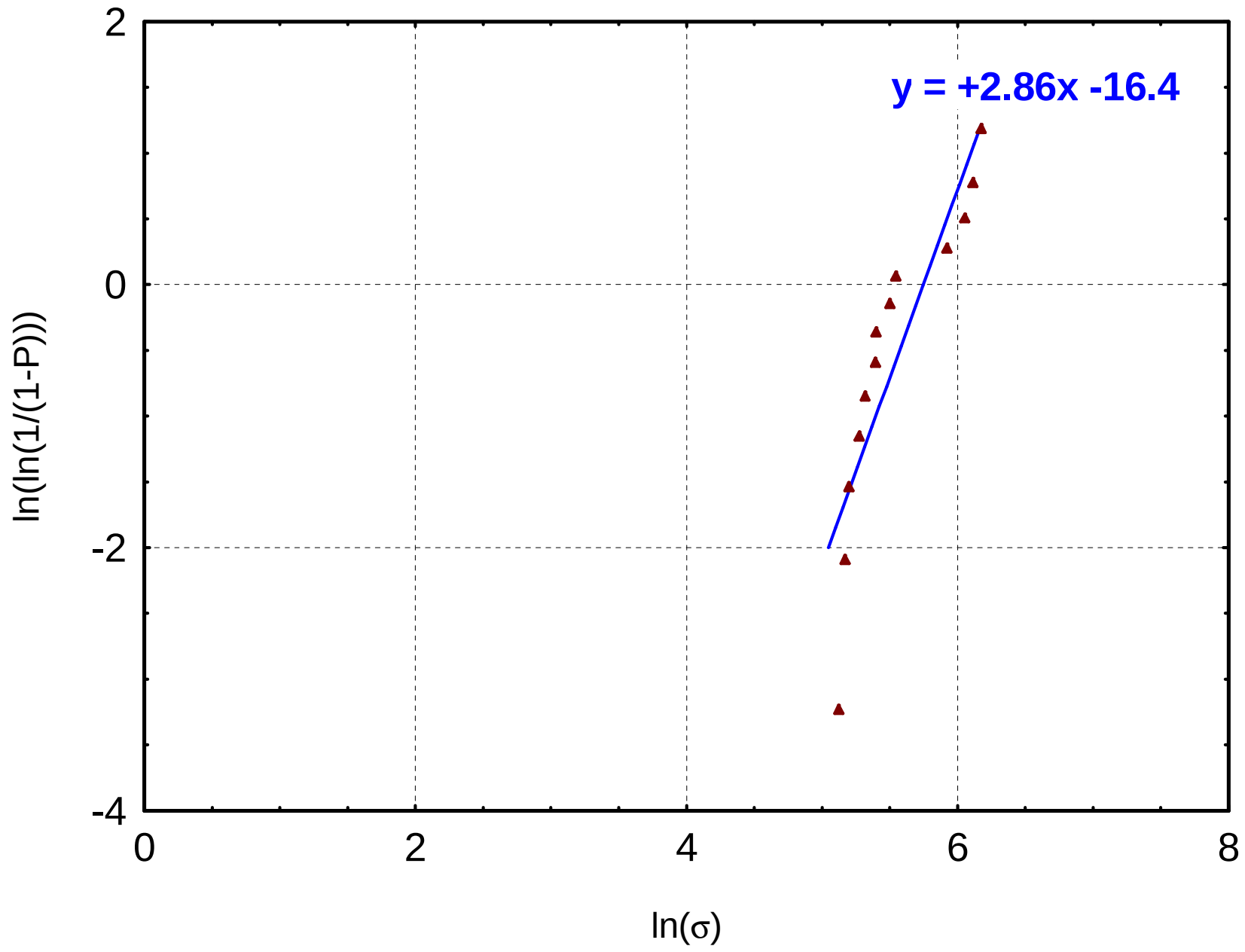

**Figure 6: Weibull distribution for hemp fibres – free length of fibre: 10 mm – 15 fibres with breaking in the central part (3 mm) of the free length**

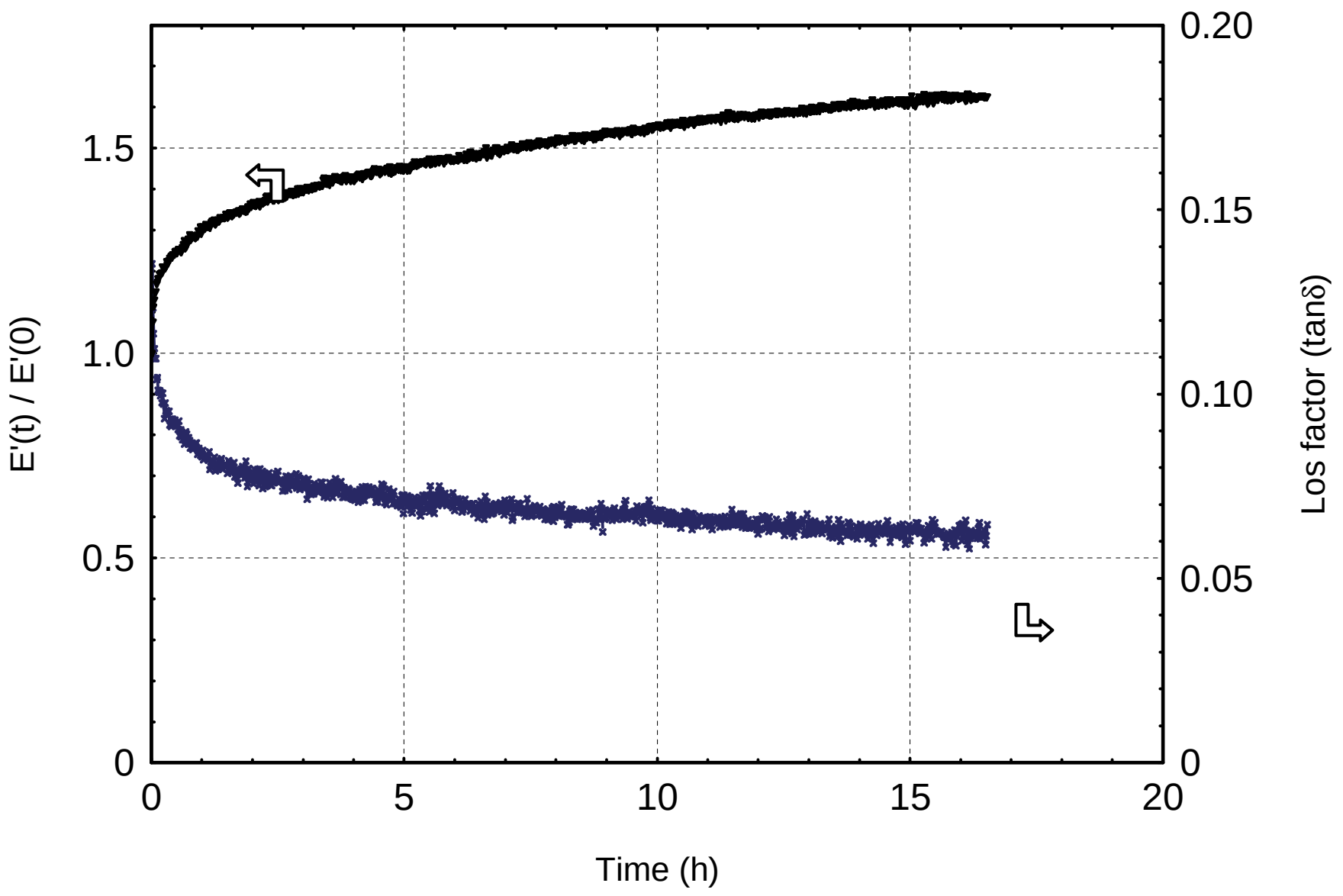

**Figure 7: Normalized storage modulus and loss factor versus time. Frequency: 1Hz. Bundle of hemp fibres. Free length: 10 mm.**

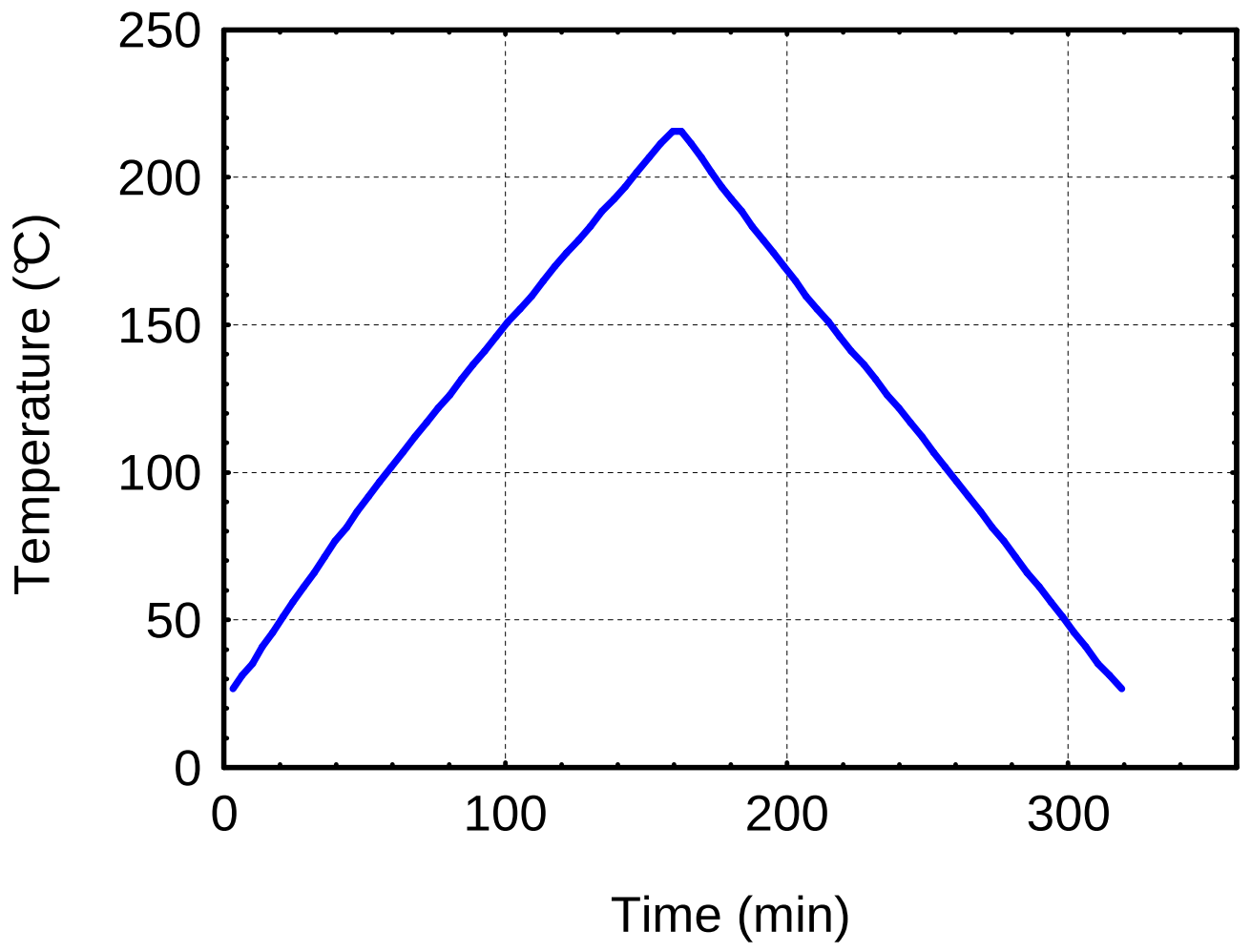

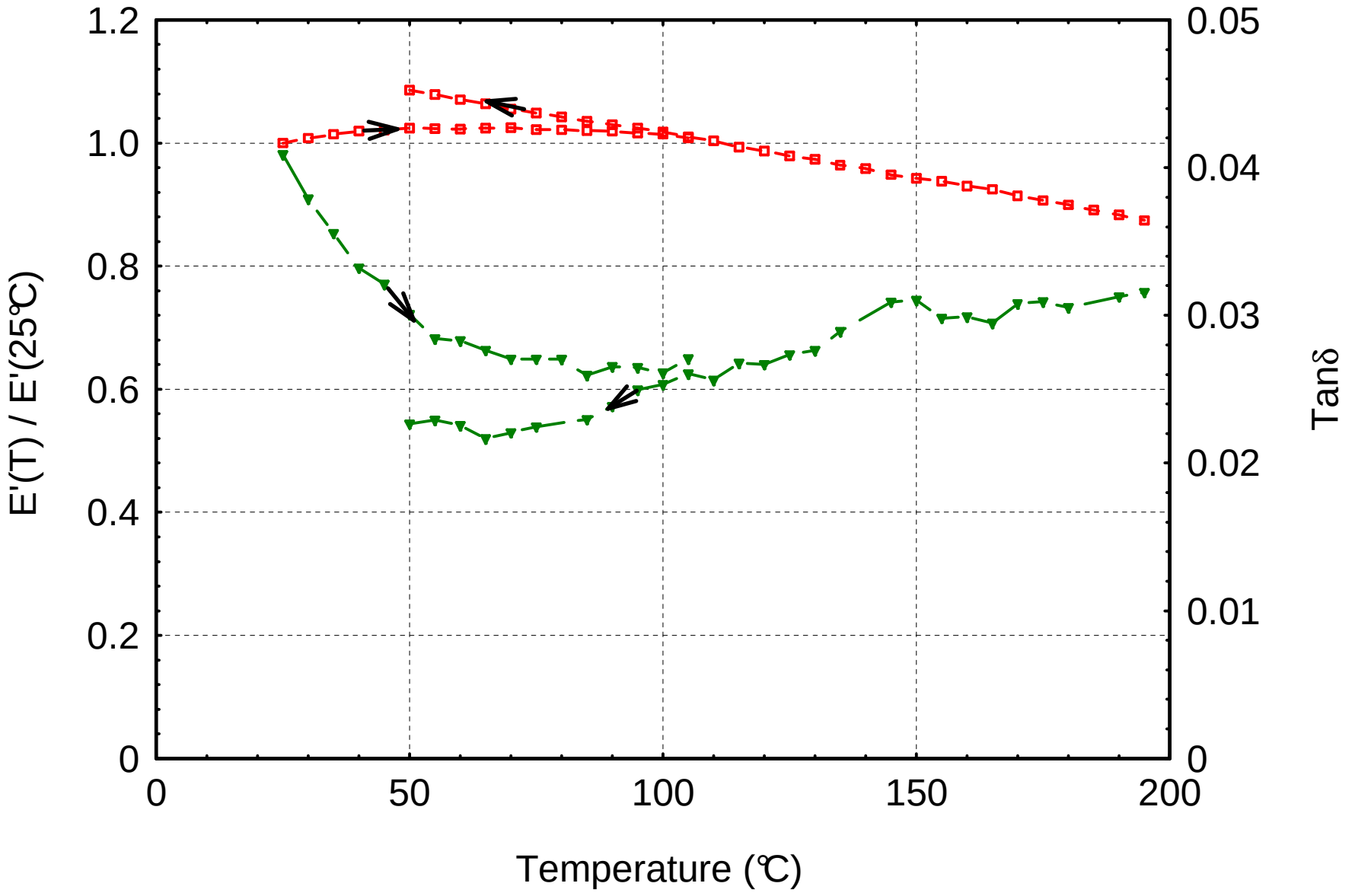

**Figure 8-a: Temperature versus time. 7-b. Normalized storage modulus and loss factor versus temperature**

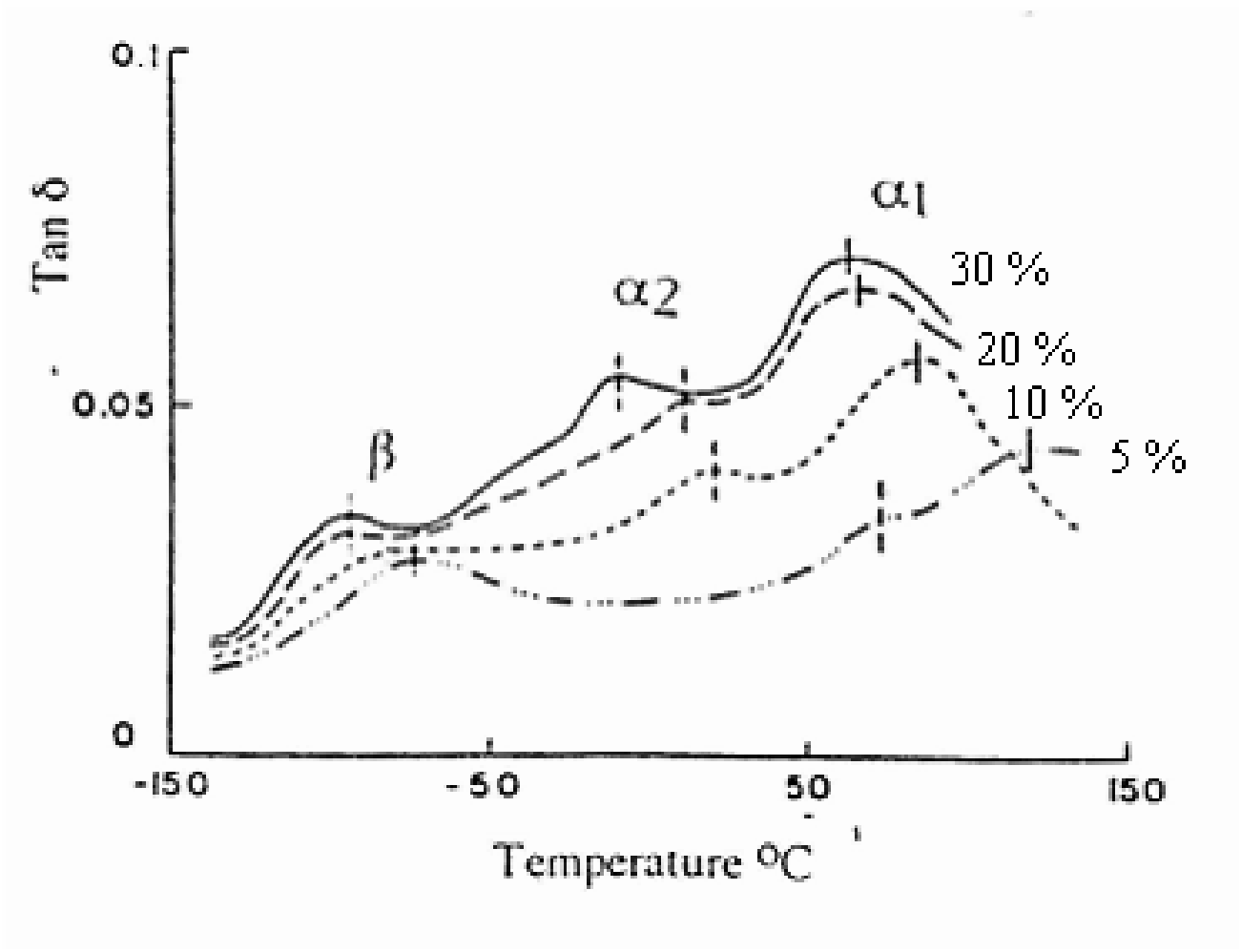

**Figure 9: Loss factor versus temperature for spruce wood at different moisture content from the original work of Kelley *et al* [16].**

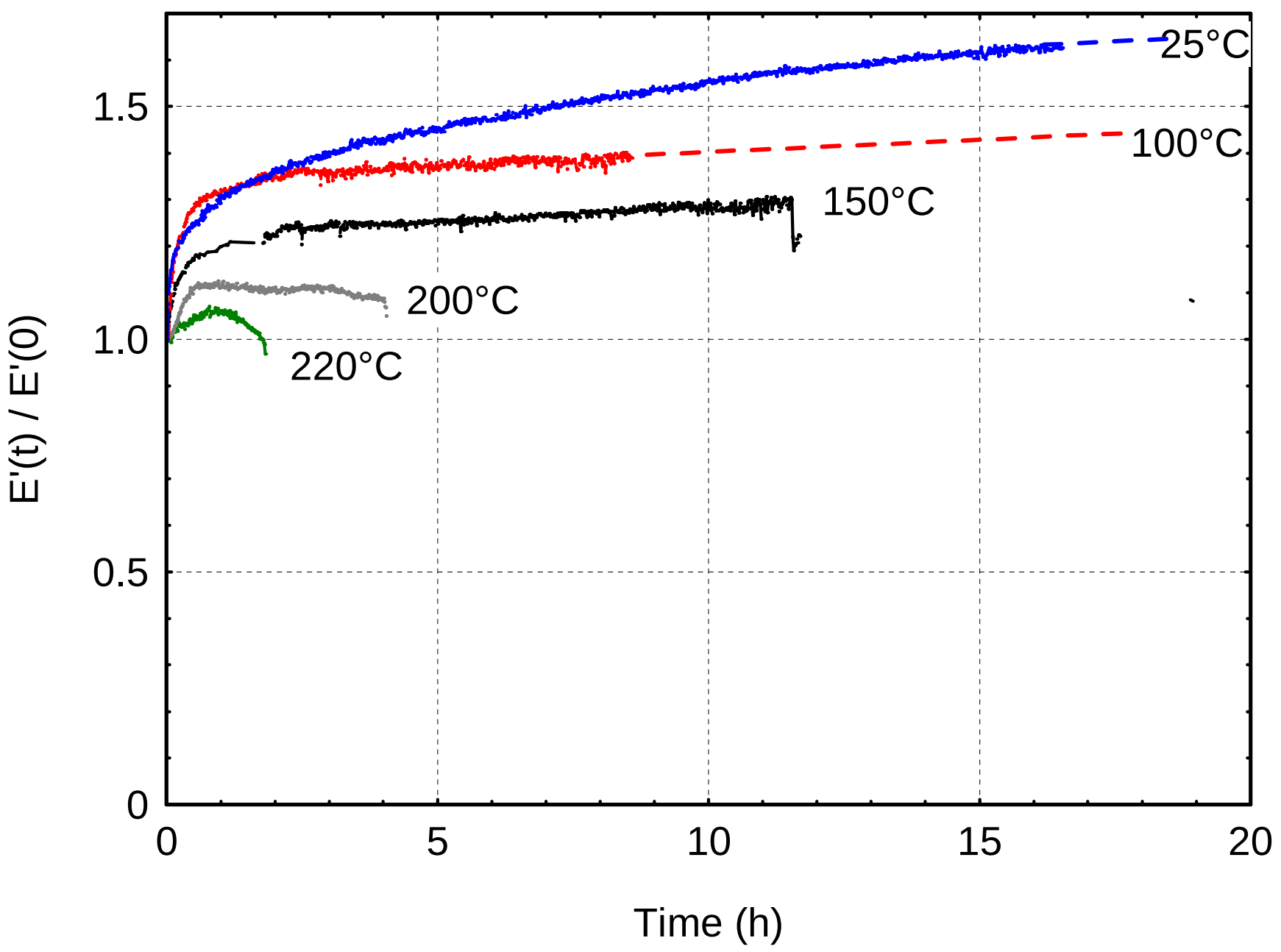

**Figure 10: Normalized storage modulus versus time at different plateau temperature**

| Fibre | | Young's modulus E (GPa) | Ultimate stress $\sigma_u$ (MPa) | Strain to failure $\varepsilon_f$ (%) | Density $\rho$ ($\kappa g.dm^{-3}$) | Specific Young's modulus E/ρ | Specific ultimate stress $\sigma_u/\rho$ |
|---|---|---|---|---|---|---|---|
| Hemp[1] | Mean | 14.4 | 285 | 2.2 | 1.07 | 13 | 260 |
| (LCDA Company) | Range | (5.6 – 30.1) | (168 – 480) | (1.3 – 3.3) | Bodros *et al*. [5] | | |
| | | | | | | | |
| Manila hemp Goda & Cao [1] | | 26.6 | 792 | / | 1.3 | 20.5 | 609 |
| Hemp Bodros *et al*. [5] | | 35 | 389 - 900 | 1.6 | 1.07 | 22 | / |
| E-Glass Goda & Cao [1] | | 55.8 – 99.6 | 1400 - 2500 | / | 2.56 | 21.8 – 38.9 | / |
| E-Glass[2] Guillon [21] | Range | 72 - 73 | 3200 - 3400 | 4.6 – 4.8 | 2.5 – 2.6 | 28 | 1270 |

**Table 1: Mechanical properties of hemp and glass fibres.**

[1]Hemp Fibre: Elementary fibre with an average diameter of 42 µm.

[2]Glass fibre: virgin filament of 10 µm in diameter.

| Temperature (°C) | | 25 | 100 | 150 | 200 | 220 |
|---|---|---|---|---|---|---|
| Number of cycles up to fracture | Mean | > 60 000 | > 60 000 | 52 782 | 15 947 | 6 560 |
| | Range | | | (41 520 – 64 043) | (6988 – 25 038) | (4 333 – 8 224) |

**Table 2: Number of cycles to failure according to temperature**

| | Young's modulus E (GPa) | Ultimate stress $\sigma_u$ (MPa) | Strain to failure $\varepsilon_f$ (%) | Density $\rho$ (κg.dm$^{-3}$) | Specific Young's modulus $E/\rho$ | Specific ultimate stress $\sigma_u/\rho$ |
|---|---|---|---|---|---|---|
| PP | 1.03 | 25 | / | 0.92 | 1.1 | 27.2 |
| PP-R hemp<br>Mf = 30% | 1.9 | 20 | 1.6 | 0.95 | 2 | 21.1 |
| PP-UD hemp<br>Mf = 30%<br>orientation 90° | 2.3 | 19 | 2.2 | 0.97 | 2.4 | 19.6 |
| PP-UD hemp<br>Vf = 30%<br>orientation 0° | 12.2 | 37 | 0.5 | 0.97 | 12.6 | 38.1 |
| | | | | | | |
| PP- R glass fibres<br>Vf = 22%<br>Baley *et al*. [4] | 6.2 | 88.6 | / | / | / | / |

*R fibres: mat of randomly scattered fibres*
*UD: mat of unidirectional fibres*
*$M_f$: mass of fibres*
*$V_f$: volume of fibres*

**Table 3: Average properties of polypropylene and manufactured composites**